\documentclass[11pt]{article}
\usepackage{amsmath,graphicx,multirow,algorithm,algpseudocode}
\usepackage{hyperref}
\usepackage{INTERSPEECH_mod}

\title{Bi-LSTM Scoring Based Similarity Measurement with Agglomerative Hierarchical Clustering (AHC) for Speaker Diarization}
\name{Siddharth S. Nijhawan$^1$ and Homayoon Beigi$^{1,2}$}
\address{
  $^1$Electrical Engineering Dept., Columbia University\\
  $^2$Recognition Technologies, Inc. and Columbia University}
\email{$^1$sn2951@columbia.edu, $^2$beigi@recotechnologies.com}

\begin{document}
%\ninept
%
\maketitle
\begin{abstract}
Majority of speech signals across different scenarios are never
available with well-defined audio segments containing only a single
speaker. A typical conversation between two speakers consists of
segments where their voices overlap, interrupt each other or halt
their speech in between multiple sentences. For a variety of
applications such as transcription, it is really important to identify
multiple speakers in a conversation, for instance, generating captions
for a discussion or a meeting. Thus, it becomes important for us to
effectively perform speaker diarization in speech signals containing
conversations among two or more speakers. Recent advancements in
diarization technology leverage neural network-based approaches to
improvise multiple subsystems of speaker diarization system comprising
of extracting segment-wise embedding features and detecting changes in
the speaker during conversation. However, to identify speaker through
clustering, models depend on methodologies like PLDA to generate
similarity measure between two extracted segments from a given
conversational audio. Since these algorithms ignore the temporal
structure of conversations, they tend to achieve a higher Diarization
Error Rate (DER), thus leading to misdetections both in terms of
speaker and change identification. Therefore, to compare similarity of
two speech segments both independently and sequentially, we propose a
Bi-directional Long Short-term Memory network for estimating the
elements present in the similarity matrix. Once the similarity matrix
is generated, Agglomerative Hierarchical Clustering (AHC) is applied
to further identify speaker segments based on thresholding. To
evaluate the performance, Diarization Error Rate (DER \%) metric is
used. The proposed model achieves a low DER of 34.80\% on a test set
of audio samples derived from ICSI Meeting Corpus as compared to
traditional PLDA based similarity measurement mechanism which achieved
a DER of 39.90\%.
\end{abstract}
\section{Introduction}
\label{sec:intro}

The process involving identification of the speaker of a particular
audio segment in a given audio file is called Speaker
Diarization~\cite{r:beigi-sr-book-2011}. In layman’s terms, speaker
diarization determines who spoke when.

Over the years, speaker diarization systems have lacked the full
utilization of advancements in deep learning techniques as compared to
speaker verification or recognition systems. Since the diarization
labels are confusing, for example, both ‘12223’ and ‘31112’ present
equally apt sequences of speaker labels throughout the audio file, and
diarization is treated as an unsupervised learning problem, there is a
need for devising a fully-supervised learning model for this problem
statement. Taking this into account, there have been recent
advancements in the use of Convolutional Neural
Networks~\cite{r-m:lukic-2016} and Recurrent-Neural
Networks~\cite{r-m:zhang-2019} for improvising the performance of
speaker diarization.

In state-of-the-art methods, PLDA is applied to estimate the
similarity metric between two speech segments, however, since PLDA is
a hypothesis testing-based method~\cite{r:beigi-sr-book-2011},
comparisons are only performed in pairs therefore completely
dismissing the time-related organization of similarity
computation. The sequential-order of speech segments is completely
disregarded due to their probabilistic nature. Since people always
converse in a structured manner and not randomly over time, using PLDA
for similarity scoring leads to a high diarization error rate. This
problem can be tackled by using a bi-directional LSTM to compute the
elements of the similarity matrix (backward as well as forward) of
audio signal. Therefore, the use of Bi-LSTM over PLDA is proposed to
compare the similarity of 2 segments both independently and
sequentially and achieve a lower diarization error rate.

In summary, the proposed work achieved the following: an end-to-end
diarization pipeline is designed using the ICSI Meeting Corpus
dataset~\cite{r-m:janin-2003} consisting of around 70 hours of meeting
recordings of multiple speakers. The Time-delayed Neural Network used
to extract x-vector embeddings is trained from scratch on a portion of
ICSI Meeting Corpus dataset and the Bi-LSTM is also trained on a
portion of the dataset through k-fold cross validation. To further
boost the performance efficiency of training Bi-LSTM and reduce the
consumption of memory, batch processing is employed which breaks down
the similarity chunks into small matrices and feeds them into the
memory sequentially. The computed DER \% based on the similarity
matrix generated from Bi-LSTM and clustering is compared with the
traditional scoring algorithm of PLDA across various parameters like
x-vector embeddings dimension, window length, AHC threshold, etc. to
showcase the low error rate of the proposed algorithm. Finally, we
also compared the performance of AHC clustering with the traditional
graph-based algorithm of Spectral Clustering (SC) for Bi-LSTM scoring.

The remaining paper is categorized as follows: Section 2 talks about
the various state-of-the-art diarization techniques along with their
implications. Section 3 formulates the problem statement through a
theoretical description of the techniques used followed by a detailed
system overview of the proposed diarization pipeline and the pseudo
code executed to achieve the results through Kaldi in Section
4. Experimental techniques and the quantitative results obtained are
detailed in the 5th Section and finally, we draw conclusions along
with scope for future work in this domain in the last section.

\section{State of the art}
\label{sec:state_of_the_art}

Typically, a lot of sub-systems are coupled together to develop a
holistic speaker diarization system. Starting with separating the
speaker audio from background noise, Voice Activity Detection
(VAD)~\cite{r-m:chen-2020,r:beigi-sr-book-2011} is performed which is
usually based on energy thresholds. Once we segregate these speech
regions from original audio, uniform segmentation~\cite{r-m:wang-2022}
is applied to further split them into segments containing
speaker-homogenous contents. In typical scenarios, this process can
also be achieved using a Speech Change Detector
(SCD)~\cite{r-m:hruz-2017,r-m:yin-2017}, which splits these speech
regions into multiple same-speaker segments. To extract features out
of these homogenous segments, a mapping to a fixed dimensional space
is applied through speaker embedding systems such as
x-vector~\cite{r-m:snyder-2018,r-m:garcia-romero-2017} or
i-vector~\cite{r-m:sell-2014}. Over the years, i-vectors have been
extensively used in the form of low-dimensional vector embeddings
computed over MFCC features~\cite{r:beigi-sr-book-2011} for automatic
speech recognition. However, while using i-vectors for speaker
diarization, a clustering layer is required as these embeddings
represent both channel as well as speaker features. Since the process
of clustering is extremely correlated with the total size of speech
segments analyzed by the system, there is a high risk of poor
performance in mapping these segments to speakers if the embeddings
process short segments of speech containing less
information~\cite{r-m:vesnicer-2014}. Due to this risk associated with
using i-vectors, anchor modelling techniques were introduced
in~\cite{r-m:mami-2002} to output a similarity score for utterance
anchors which represent the speech utterances from a set of
pre-trained speak models. Several diarization algorithms also employ
speaker verification methods~\cite{r-m:cai-2018,r-m:cai-2018-1} to
generate these feature embeddings from the outputs present in the
penultimate layer. \cite{r-m:rouvier-2015} performed speaker
classification by training a 3-layer neural network which was then
applied to a Gaussian Mixture Universal Background model. As a scoring
mechanism, various similarity measurement techniques like
Probabilistic Linear Discriminant Analysis
(PLDA)~\cite{r-m:prince-2007} or Cosine Similarity are used to
identify similarity metric between a pair of these segments to
generate a similarity matrix. To obtain diarization results,
similarity matrix is passed on as an input to various clustering
algorithms like Spectral Clustering~\cite{r-m:wang-2022},
Agglomerative Hierarchical Clustering (AHC)~\cite{r-m:meignier-2010},
etc.

\section{Problem Formulation}
\label{sec:problem_formulation}

A typical speaker diarization system is tasked with the objective of
identifying the set of labels depicting when each speaker talks by
analyzing a given set of speech signals. In terms of computational
learning paradigm, we can formulate this problem as a typical
supervised learning-based classification task provided we know the
identities of speakers in some form of the data input to the
system. However, this is an extremely ideal case, and does not occur
in real world. So, to approach the problem of speaker diarization, we
can split it into two stages.

Firstly, we classify each of the speakers by training a time-delayed
neural network which can extract time-dependent speaker
characteristics or speaker embeddings called
x-vectors~\cite{r-m:snyder-2018}. Generally, the activations generated
from the penultimate layer of the neural network are used as
x-vectors. These x-vectors are obtained by aggregating the outputs
after sigmoid layer in a class-by-class manner followed by normalizing
these values over the entire audio signal.

Once we have extracted all the speaker dependent information, we
analyze these embeddings as a function of time so that the
computational algorithm can detect when the speaker changes. The
speakers which are new to the system are then compared with the
existing database of previous speakers’ feature embeddings through a
similarity measurement methodology. In layman’s terms, if similarity
measure between two embeddings is below a particular user-defined
threshold, the speaker is considered as new, otherwise it is mapped
with the closest speaker embedding. This process can also be
implemented using a learning algorithm (Bi-LSTM in our case) where the
primary objective of the neural network is to predict elements of
similarity matrix between each of the speaker embeddings. For this
supervised learning task, we input the speaker embeddings as the
features and the ground truth labels based on speaker identity
information. Equation~\ref{eq1} denotes the Binary Cross Entropy (BCE)
loss function which the neural network aims to optimize for N training
samples and n classes (dimension of similarity matrix).

\begin{equation} \label{eq1}
L(y, \Bar{y}) = -\sum_{a=1}^{N}\sum_{b=1}^{N}y_a^blog(\Bar{y}_a^b)
\end{equation}

The predicted similarity measure is denoted by \(\Bar{y}_a\) for \(a_{th}\) data point and \(\Bar{y}_a\) denotes the ground truth for the same data point. 

\subsection{x-vector embeddings}
\label{ssec:x-vector_embeddings}

x-vector is a type of feature embeddings extracted using deep neural
network which was originally used in speaker verification systems as
features. These x-vectors are obtained through supervised learning of
a time-delayed neural network where MFCCs extracted from the speech
data are used as input features. The various frame-level features are
transformed into a single segment-level embedding through time-pooling
modules of time-delay neural network. x-vector is the output of the
second last layer in the neural network.

For the given problem statement of speaker diarization,
Table~\ref{table1} provides a holistic view of neural network
architecture which is trained as an x-vector extractor.

\begin{table}[h]
\small
\begin{center}
\begin{tabular}{|l|l|c|c|} 
\hline
\bf{\scriptsize Name} & \bf{\scriptsize Layer Type} & \bf{\scriptsize Input Size} & \bf{\scriptsize Output Size} \\
\hline
\hline
\bf{\scriptsize tdnn1} & {\scriptsize relu-batchnorm-layer} & {\scriptsize 13} & {\scriptsize 512} \\ 
\hline
\bf{\scriptsize tdnn2} & {\scriptsize relu-batchnorm-layer} & {\scriptsize 1536} & {\scriptsize 512} \\ 
\hline
\bf{\scriptsize tdnn3} & {\scriptsize relu-batchnorm-layer} & {\scriptsize 1536} & {\scriptsize 512} \\ 
\hline
\bf{\scriptsize tdnn4} & {\scriptsize relu-batchnorm-layer} & {\scriptsize 512} & {\scriptsize 512} \\ 
\hline
\bf{\scriptsize tdnn5} & {\scriptsize relu-batchnorm-layer} & {\scriptsize 512} & {\scriptsize 1500} \\ 
\hline
\bf{\scriptsize stats} & {\scriptsize stats-layer (pooling)} & {\scriptsize 1500T} & {\scriptsize 3000} \\ 
\hline
\bf{\scriptsize tdnn6} & {\scriptsize relu-batchnorm-layer} & {\scriptsize 3000} & {\scriptsize 512 or 128} \\ 
\hline
\bf{\scriptsize tdnn7} & {\scriptsize relu-batchnorm-layer} & {\scriptsize 512 or 128} & {\scriptsize 512} \\ 
\hline
\bf{\scriptsize output} & {\scriptsize output-layer} & {\scriptsize 512} & {\scriptsize N} \\ 
\hline
\end{tabular}
\caption{Architecture design for x-vector extractor}
\label{table1}
\end{center}
\end{table}%

Here, T is the number of frames present in the input and N represents
the number of speakers in the training set. Layers tdnn1-5 correspond
to feature-level layers in the speech containing a small context
centered around the frame currently in processing. ‘stats’ layer or
statistics pooling layer computes the mean and standard deviation
after adding all the T frame-level outputs from previous layer
tdnn5. The output of stats layer contains a 1500-dimensional vector
for each input segment T. Further, segment level layers comprising of
tdnn6 and tdnn7 aggregate the computed mean and standard deviation to
the output layer containing a SoftMax operation with the number of
identifiable speakers as the output size (class size). The size of
penultimate layer tdnn6 (or the affine component of tdnn6) determines
the dimension of x-vector embeddings which is 512 or 128 depending on
the experiments performed in the later sections.

\subsection{PLDA based Similarity Measurement}
\label{ssec:plda}

Probabilistic Linear Discriminant Analysis or PLDA is a
state-of-the-art algorithm used for computing similarity scores
between any two segments of speech (or any other form of data). Once a
PLDA system is trained on a given set of features (x-vector embeddings
in our case), hypothesis testing is used to compute similarity between
2 segments, say, a and b as described in equation Equation~\ref{eq2}.

\begin{equation} \label{eq2}
S_{ab} = F_{plda} (x_a, x_b)
\end{equation}

Here, \(S_{ab}\) is the similarity measurement between \(x_a\) and
\(x_b\). PLDA originally outputs a score between [-1, 1] which is not
ideal for clustering. Therefore, we normalize the output score using a
logistic function to bound the similarity measure between [0, 1]. The
logistic function l(x) is defined in Equation~\ref{eq3}.

\begin{equation} \label{eq3}
l(x) = \frac{1}{1 + e^{-5x}}
\end{equation}

Therefore, now \(S_{ab}\) is bounded between [0, 1] where 1 denotes
that segments a and b originate from a single speaker and 0 denotes
otherwise.

\subsection{Bi-LSTM based Similarity Measurement}
\label{ssec:bi-lstm}

An ideal similarity matrix contains Boolean elements where 0 denotes
no similarity and 1 denotes that the two elements are from the same
speaker. Moreover, content of the matrix does not change with the
change in speaker index. To treat this problem as a supervised
learning problem, the entire speaker embedding sequence x is used with
matrix S as the class label. This is how we formulate the objective
for Bi-LSTM model optimization. Therefore, we use binary cross entropy
loss during the training of Bi-LSTM model to predict each row of S.

First step is to concatenate 2 x-vectors \(x_a\) and \(x_b\) which
generates a 2D input for LSTM in the form \([x_a^T, x_b^T]^T\) having
the output as \(S_{ab}\). Equations~\ref{eq4} and~\ref{eq5} depicts
the formulation of learning problem for Bi-LSTM in sequential manner.

\begin{equation} \label{eq4}
S_a = [S_{a1}, S_{a2}, . . ., S_{an}]
\end{equation}

\begin{equation} \label{eq5}
[S_{a1}, S_{a2}, . . ., S_{an}] = F_{bilstm}(\begin{bmatrix} x_a\\x_1\end{bmatrix}, \begin{bmatrix} x_a\\x_2\end{bmatrix}, . . ., \begin{bmatrix} x_a\\x_n\end{bmatrix})
\end{equation}

\begin{figure}[htp]
\begin{minipage}[b]{1.0\linewidth}
    \centering
    \includegraphics[width=3in]{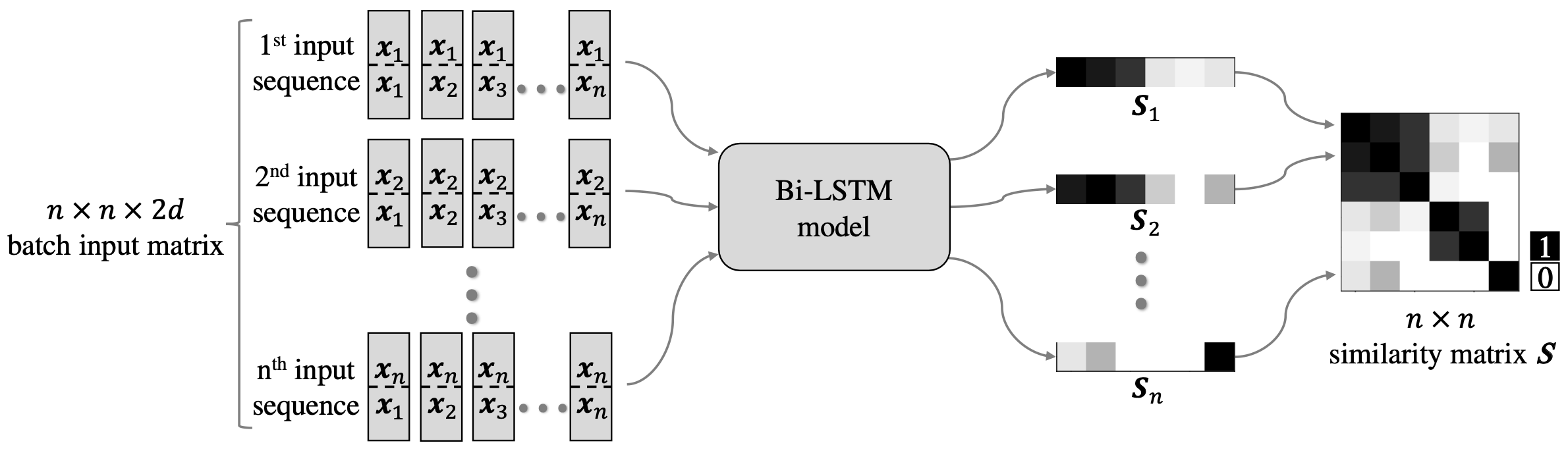}
    \caption{Bi-LSTM System Design~\cite{r-m:lin-2019}}
    \label{figure1}
\end{minipage}
\end{figure}

Here, \(S_a\) also depicts the output of \(a^{th}\) sequence in a
batch containing a total of n sequences. Therefore, to form the
similarity matrix S, each of the n outputs are stacked
row-wise. Figure~\ref{figure1} shows the high-level architectural
working of Bi-LSTM based similarity measurement.

\vfill\pagebreak

For audio signals, the value of n is usually large leading to the size
of matrix S being extremely large. For instance, if n equals 10,000
and d equals 512, the total size of batch input matrix will be 10,000
x 10,000 x (2 x 512), i.e., 1024 x \(10^{8}\). If each data point is stored
as a floating-point datatype (requiring 4 bytes of memory), the matrix
will require around 190.73 GB of RAM to perform computations on the
entire matrix at once. Apart from this, LSTMs usually have poor
generalization performance when given very long sequences as an
input. The challenge of memory requirement can be solved by using the
technique of sliding window, however, the similarity matrix generated
will be of the form of a diagonal block. This will lead to the system
being unable to identify the different or same speakers among
different windows.

\begin{figure}[htp]
    \centering
    \includegraphics[width=3in]{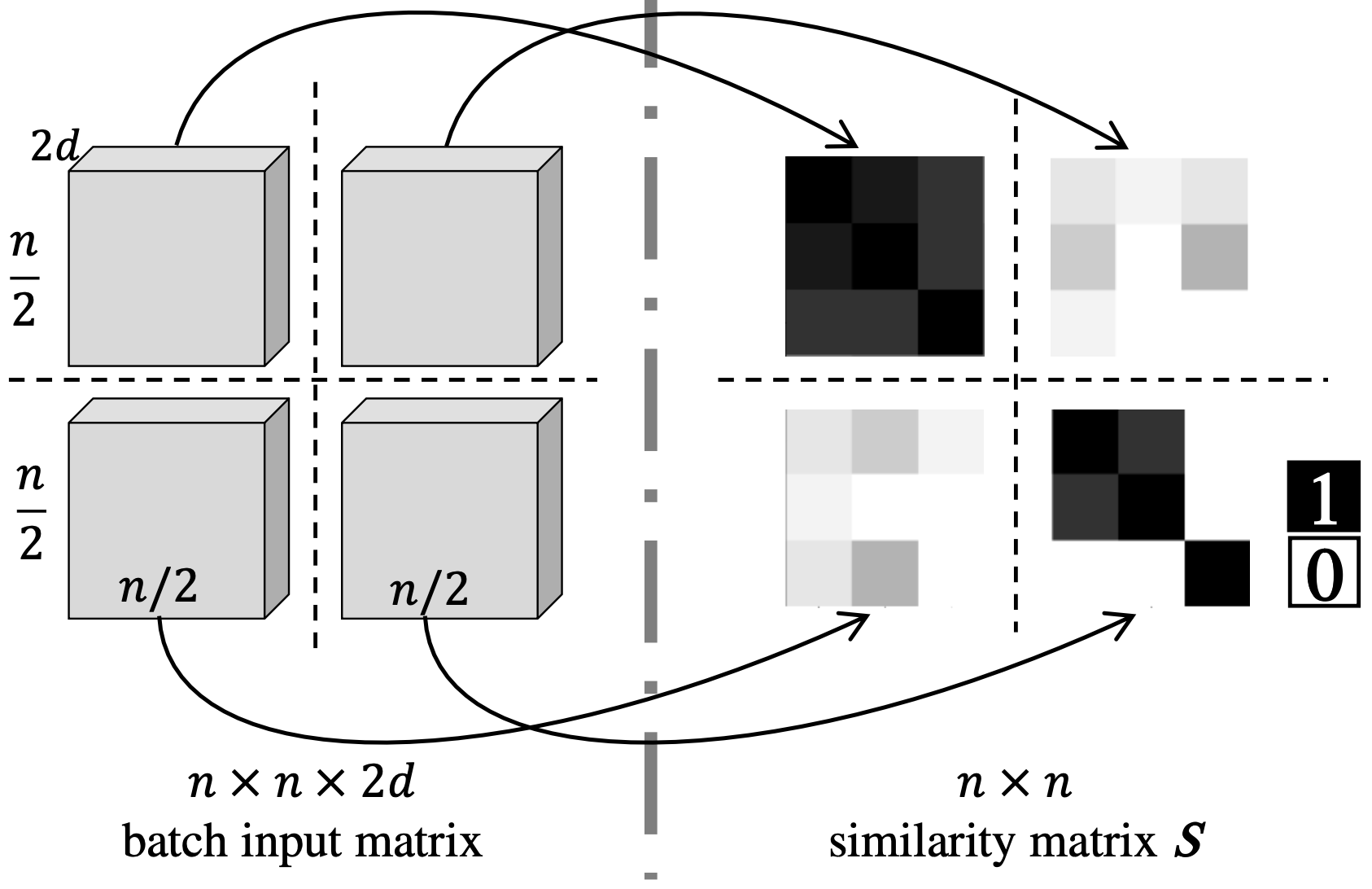}
    \caption{Batch-processing technique~\cite{r-m:lin-2019}}
    \label{figure2}
\end{figure}

In the proposed work, we suggest the technique of batch processing to
tackle the above-mentioned shortcomings. Similarity matrix S is
divided into several small chunks of matrices with size dependent on a
max length threshold and process these batches sequentially through
the Bi-LSTM model. Figure~\ref{figure2} denotes the breakage of a
single \(n \times n\) matrix into 4 sub-matrices of size \(\frac{n}{2}
\times \frac{n}{2}\) which is then passed onto the Bi-LSTM network
sequentially.

In terms of neural network architecture, the model consists of 2
Bi-LSTM layers having 512 outputs each. Since the LSTM is
bidirectional, 256 outputs are in forward direction and 256 are in
backwards direction. This is followed by a fully connected layer
having 64 dimensions and ReLU activation layer. The final layer is a
single dimensional layer connected to a Sigmoid operation. The Sigmoid
function is responsible for generating the similarity measurement
between 0 and 1.

\subsection{Spectral Clustering (SC)}
\label{ssec:spectral_clustering}

Spectral Clustering (SC) is a clustering algorithm which is based on
graphs~\cite{r-m:vonluxburg-2007}. To compute values of a similarity
matrix S, SC generates an undirected graph with the number of nodes
equal to the number of rows or columns in S. All the nodes are
connected with edges having weights equal to \(S_{ab}\) (for edge
between a and b). SC then removes edges with weights less than a
threshold value and hence, forms multiple sub graphs from the existing
graph.

As a first step in SC, every single diagonal element is set as 0
because it denotes self-similarity. Then Laplacian matrix L is
formulated using the difference between diagonal matrix D defined as
\(D_a = \sum_{b=1}^{n}S_{ab}\) and similarity matrix S (as shown in
Equation~\ref{eq6}).

\begin{equation} \label{eq6}
L = D - S
\end{equation}

Here, the norm of Laplacian matrix is computed in equation~\ref{eq7}.

\begin{equation} \label{eq7}
L_{norm} = D^{-1}L
\end{equation}

After computing eigenvalues and eigenvectors of \(L_{norm}\). SC then
takes k smallest eigenvalues and their corresponding eigenvectors to
construct a matrix P containing each column as the set of k smallest
eigenvectors. Finally, each row of P is clustered using k-means to
generate the similarity matrix.

\subsection{Agglomerative Hierarchical Clustering (AHC)}
\label{ssec:ahc}

Agglomerative Hierarchical Clustering (AHC) is a form of hierarchical
clustering methodology where the objective of the algorithm is to
perform consecutive unification
operations~\cite{r-m:gowda-1978}. Unification or merging occurs when
two particular data points are assigned the same cluster based on a
similarity measure. These similar clusters are then further used for
clustering. AHC algorithm starts by initializing clusters equal to the
total number of datapoints, n in our case (number of rows or columns
of similarity matrix \(S_{ab}\)). As the next step, algorithm looks
for the pair having highest similarity, unifies them, and subtracts 1
from the total number of available clusters. This process is
recursively repeated with the stopping condition that the similarity
measure between any 2 clusters falls below a particular value
designated by the user.

\section{Proposed Methodology}
\label{sec:proposed_methodology}

\begin{figure*}[htp]
  \centering
  \begin{minipage}[b]{1.0\linewidth}
    \begin{center}
    \includegraphics[width=6in]{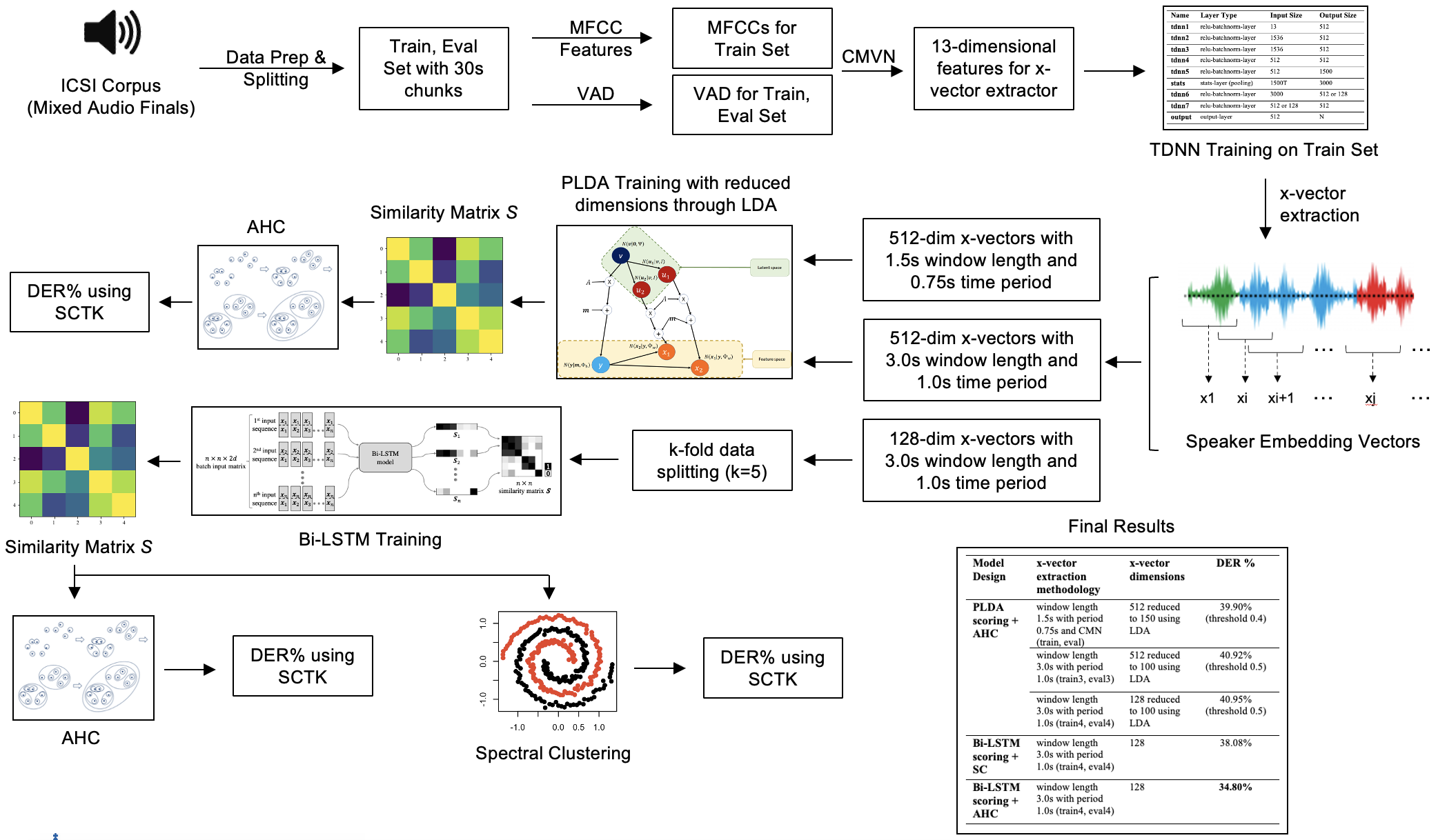}
    \caption{System Architecture}
    \label{figure3}
    \end{center}
  \end{minipage}
\end{figure*}

Figure~\ref{figure3} describes a high-level flow of the proposed
diarization model. From a given set of audio signals (obtained from
ICSI Corpus) containing both voice and background noise, we first
prepare the data for processing by splitting it into train and eval
directories (~93\% \& ~4\% respectively) followed by splitting each
speaker’s data into 30-second chunks. This step ensures a baseline
form of diarization which will assist in the further process of actual
diarization. From these 30s chunks, we extract MFCCs and perform Voice
Activity Detection to generate speech separated audio
signals. Finally, to generate input features for time-delayed neural
network, cepstral mean and variance normalization is performed to
generate set of 13-dimensional audio features for both train and eval
sets.\\

%figure* makes the figure span the whole page

For developing the x-vector extractor, a time-delay neural network is
trained on the train set of the prepared data to generate feature
embeddings in the form of x-vector \([x_1, x_2, .... x_n]\). As
mentioned in the previous section, 2 different sets of neural networks
are trained: one to generate 512-dimensional x-vectors and another to
generate 128-dimensional x-vectors. Once the training is complete, a
set of x-vectors are extracted for both train and eval sets using
varying window size and time period.

A bi-directional LSTM model is used to predict similarity score
\(S_a\) of every embedding vector pair (\(x_a\), \(x_b\)) to generate
similarity matrix S using the binary cross entropy loss function. To
compare the performance of our model with state-of-the-art diarization
techniques, a PLDA model is also trained on the train set x-vectors
after reducing the dimensions through LDA (150-dim and 100-dim for
experimentation). These reduced set of x-vector features are then
scored on eval set features to generate similarity matrix.

For the task of supervised diarization learning, we leverage the
entire matrix S as the class for the given speaker embedding
sequence. Once similarity matrix is generated, Agglomerative
Hierarchical Clustering is applied which initializes each segment as a
singleton cluster. Since AHC algorithm is represented as a binary-tree
building process, it works from bottom to top by considering each
cluster as a leaf. During learning iterations, we merge clusters
having a large similarity value and stop when the score is below a
particular threshold hyperparameter value. This process is repeated
for similarity matrix generated by PLDA as well. Again, we also apply
Spectral Clustering algorithm for both Bi-LSTM and PLDA based
similarity matrices. The final Diarization Error Rate (DER\%) are
generated by using Statistical Language Modelling Toolkit (SCTK) which
compares the generated segment labels with ground truth present in
respective utt2spk files.

Kaldi speech recognition toolkit~\cite{r-m:povey-2011} is used to
create the end-to-end diarization pipeline. The pseudo code for the
entire process is depicted by Algorithm 1.

% \vfill\pagebreak

% \begin{algorithm}
% \caption{Diarization pipeline in Kaldi}\label{algo1}
% \begin{algorithmic}
% \Require $n \geq 0$ \\
% \textbf{Input:}
% \Ensure $y = x^n$
% \end{algorithmic}
% \end{algorithm}

\begin{table}[htb]
\centering
\begin{minipage}[b]{1.0\linewidth}
\begin{center}
% \caption{Architecture design for x-vector extractor}
\begin{tabular}{ p{0.9\linewidth} } 
\textbf{Algorithm 1: Diarization Pipeline in Kaldi} \\
\hline

\textbf{Input:} ICSI Corpus Dataset (individual headset mix) + Transcriptions \\
\\
\textbf{\textit{run\_prepare\_shared.sh}} \\
-prepare dictionary ./icsi\_prepare\_dict.sh \\
-prepare language resources ./prepare\_lang.sh \\
-convert transcriptions from MRT format to annotations ./icsi\_text\_prep.sh \\
-generate language model ./icsi\_train\_lms.sh \\
\\
\textbf{\textit{run.sh}} \\
-split data directories: dev, train, eval \\
-for data in train dev eval... do: \\
    ./modify\_speaker\_info.sh 30 (split 30s chunks) \\
-mfcc extraction for train set ./make\_mfcc.sh \\
-for data in train dev eval... do: \\
    ./compute\_vad\_decision.sh \\
-for data in train dev eval... do: \\
    ./prepare\_feats.sh
\\
-for d = 512 and 128... train x-vector extractor ./run\_xvector\_1a.sh \\
-for d = 512 and 128... extract x-vectors ./extract\_xvectors.sh \\
-for d = 512 and 128... train and score using plda ivector-compute-plda \\
-for d = 512 and 128... predict DER using AHC ./cluster.sh \\
\\
-for d = 128... 5-fold split ./ kfold.py \\
-for each split... train bi-lstm ./train.py \\
-for each split... predict similarity matrix ./predict.py \\
-for each split... compute DER using AHC and SC ./cluster.py \\
\\
\textbf{Output:} DER logs per experiment for eval set\\

\hline
\label{algo1}
\end{tabular}
\end{center}
\end{minipage}
\end{table}%

\subsection{Dataset}
\label{ssec:dataset}

The proposed work utilizes ICSI Corpus [11] for training various
methodologies in the pipeline as well as evaluation. ICSI Corpus is a
dataset which is based on multi-channel audio samples extracted from a
set of 75 meetings entirely based in English language. These meetings
have been collected between the time period 2000-2002 which occurred
at International Computer Science Institute, Berkeley. The minimum
length of a meeting is 17 minutes whereas meetings as long as 103
minutes are present in the dataset as well. In total, around 72 hours
of audio data is present in the form of meeting room speech. The
dataset also contains transcriptions for each of these meetings along
with specific annotations of non-speech as well as speech segments
within the data. This information is present in the MRT extension. In
terms of demographical information, each meeting contains around 3 to
10 participants, with a total of 53 unique speakers in the entire
dataset. To add to the variation, dataset also contains a good amount
of non-native English speakers having different levels of fluency. For
our pipeline, the speech segments are divided into 3 sets of training
(67.5 hours), development (2.2 hours) and evaluation (2.8 hours) which
ensures that there is a minimum overlap of same speakers across these
sets. Dataset was originally recorded in 3 different types: individual
headset mic recording, distant multiple mics recording and distant
single mic recording. For our purpose, we used individual headset
recordings. The information regarding the speaker mapping to the
headset is present in the MRT transcription files which is decoded
through Kaldi.

\subsection{Implementation Details}
\label{ssec:implementation_details}

As the entire pipeline is designed using Kaldi, there exists a recipe
to prepare the ICSI Corpus for individual headset mic recording
type. First, we execute the script run\_prepare\_shared.sh available
in Kaldi’s ICSI Corpus recipe for ASR to generate dictionary, language
resources, language model and annotations. Post this step, everything
is implemented in the script titled run.sh which is designed from
scratch and suited to ICSI Corpus dataset.

In the initial stages, we do some data-preprocessing: splitting data
into dev, test, and train sets, and making 30s chunks of speaker
speech segments. Kaldi’s pre-designed bash scripts are used for these
stages. Once the data is ready, MFCCs are extracted, followed by VAD
and then generating final set of 13-dimensional features post
CMVN. For our experiments involving PLDA, we train an x-vector
extractor from the scripts present in SRE16 recipes of
Kaldi. run\_xvector\_1a.sh script is executed which generates examples
for training the TDNN and finally, trains the TDNN. For this script,
we hardcode min-frames-per-chunk as 16 and max-frames-per-chunk as
50. With these examples, we train 2 TDNN’s (512 and 128-dimensional
x-vector extractors). For our experiments involving PLDA, we use
Kaldi’s in-built scripts ivector-compute-lda, ivector-compute-plda,
score\_plda.sh to generate PLDA-based scores for window length and
time period as discussed in the experiments section. Next, AHC is
applied using cluster.sh script included with Kaldi where we search
for the optimum threshold between -0.3 and 0.5. To evaluate the DER,
we use SCTK’s md-eval.pl script to generate diarization report for
each of these experiments.

Bi-LSTM network is trained using a readily available library for
Kaldi~\cite{r-m:lin-2019} which contains script to train a custom
Bi-LSTM network, split the data to perform k-fold validation, and
clustering (AHC and SC). We train the Bi-LSTM for a total of 10 epochs
due to constraints in time and resources. For splitting the data into
batches, we used a threshold of 200, i.e., if sequence length is above
200, it will be broken down into batches dependent on the size of
sequence. Learning rate is set as 0.01 initially in the training
process. Once, the network is trained, we perform AHC and SC
clustering with varying thresholds between 0 and 1.0 to find the best
optimum value for each fold. These results per fold are then combined
to generate predictions using predict.py script. Finally, cluster.py
script uses SCTK’s md-eval.pl to generate DER results for best
possible threshold value.

\section{Results and discussion}
\label{sec:results}

\subsection{Evaluation Metric: Diarization Error Rate (DER\%)}
\label{ssec:der}

DER is a metric used for quantitatively evaluating speaker diarization
modules. The following errors are included in computing the DER:
errors from voice activity detection, segmentation error, and
classification error. DER \% can be computed with the expression
depicted in Equation~\ref{eq8}.
 
\begin{equation} \label{eq8}
DER\% = \frac{err_{spk} + err_{fas} + err_{miss}}{T} \times 100 
\end{equation}
 
The numerator is an absolute sum of total time incorrectly identified
as voice (\(err_{spk}\)), total timestamps assigned to incorrect
speakers (\(err_{fas}\)), and the amount of speech missed due to
faulty voice activity detection (\(err_{miss}\)). Here, T denotes the
total time of the audio sequence. Since the annotation of speech data
is performed manually, there is always a chance for human error. DER
computation also takes into account the possibility of human error by
providing an acceptance margin of 250ms.
 
\subsection{Quantitative Evaluation}
\label{ssec:quantitative_evaluation}

Quantitative evaluation is performed by carrying out experiments to
determine DER\% for two different scoring algorithms: PLDA and
Bi-LSTM. For PLDA based scoring, to demonstrate the trade-off between
memory consumption and accuracy, we use 2 different sets of x-vector
models having 512 and 128 dimensions respectively. Also, 2 different
types of window lengths and time periods are used to generate separate
sets of x-vectors which, however, did not have a noticeable impact on
the accuracy of model. Finally, to test the efficacy of our proposed
pipeline using Bi-LSTM scoring and AHC clustering, we also tested the
model by using Spectral Clustering to generate DER for eval
set. 128-dimensional x-vector extractor was used for both of the
experiments involving Bi-LSTM with a window length of 3.0s having time
period 1.0s during extraction. Apart from the first experiment
containing 512-dimensional x-vectors on PLDA + AHC pipeline, the
sliding window cepstral mean normalization was not applied.

Table~\ref{table2} contains the resulting DER \% on the eval set of
ICSI Corpus for different model designs and experiments. As evident
from the table, the proposed algorithm which uses Bi-LSTM network for
scoring and AHC for clustering achieved the least DER of 34.80\% which
is a noticeable improvement in comparison to the DER of range 39.9\% -
43.51\% achieved by state-of-the-art PLDA scoring and AHC based
system. Since LSTMs have the capability to learn the sequential
patterns in the data, our proposed work performed better in the
classification task of identifying speakers for each of the audio
segments. PLDA does not take into account the sequential information
of conversation, i.e., how speakers take turns in talking and perform
a highly structured conversation. Therefore, as evident from our
analysis, Bi-LSTM is able to fully understand the statistical
information in conversations with the help of its forward and backward
layers.

\begin{table}[h]
\footnotesize
\begin{center}
\begin{tabular}{ p{0.2\linewidth} p{0.25\linewidth} p{0.15\linewidth} p{0.15\linewidth} } 
\hline
\textbf{Model Design} & \textbf{x-vector extraction methodology} & \textbf{x-vector dimensions} & \textbf{DER\%} \\
\hline
\textbf{PLDA scoring + AHC} & window length 1.5s with period 0.75s and CMN & 512 reduced to 150 using LDA & 39.90\% (threshold 0.4) \\ 
\cline{2-4}
& window length 3.0s with period 1.0s and no CMN & 512 reduced to 128 using LDA & 43.51\% (threshold 0.5) \\ 
\cline{2-4}
& window length 3.0s with period 1.0s and no CMN & 512 reduced to 100 using LDA & 40.92\% (threshold 0.5) \\ 
\cline{2-4}
& window length 3.0s with period 1.0s and no CMN & 128 & 43.26\% (threshold 0.5) \\ 
\cline{2-4}
& window length 3.0s with period 1.0s and no CMN & 128 reduced to 100 using LDA & 40.95\% (threshold 0.5) \\ 
\hline
\textbf{Bi-LSTM scoring + SC} & window length 3.0s with period 1.0s and no CMN & 128 & 38.08\% \\ 
\hline
\textbf{Bi-LSTM scoring + AHC} & window length 3.0s with period 1.0s and no CMN & 128 & \textbf{34.80}\% \\ 
\hline
\end{tabular}
\caption{DER \% for different model designs}
\label{table2}
\end{center}
\end{table}%

For PLDA based similarity matrix generation pipeline, reducing the
number of dimensions for x-vectors also increased the DER with a
trade-off between memory consumption and accuracy of
classification. As the dimensions are reduced from 150 to 100, the DER
drops from 39.9\% to 40.92\%. However, there is a significant drop in
the memory requirement as the total datapoints reduce from \(n \times
n \times (2\times150)\) to \(n\times n \times (2\times 100)\).

In our experiments, AHC is proven to perform better than SC which
achieved a DER of 38.08\% for Bi-LSTM based similarity matrix as
compared to 34.80\% for AHC.

\section{Conclusion and future scope}
\label{sec:conclusion}

In conclusion, we proposed an alternative speaker diarization pipeline
which leverages the sequential property of Bi-LSTM to predict
similarity measures between two datapoints instead of state-of-the-art
PLDA based scoring algorithm. The training as well as evaluation of
the computational learning algorithms like TDNN, LSTM, PLDA, etc. was
performed on ICSI Corpus which contains around 72 hours of human
speech in the form of meeting conversations. We also demonstrated the
trade-off between memory consumption and accuracy in terms of DER and
proposed a batch-processing methodology to train the extremely deep
Bi-LSTM. We performed experiments to choose the best clustering
methodology for generating the final output and chose AHC over
Spectral Clustering based on its improved performance. Our best
performing model achieved a low DER of 34.08\% on the evaluation set
extracted from ICSI Corpus.

In future, we plan to expand this work by performing data augmentation
on the dataset which adds noise, reverberation, babble noises, and
music to the original audio files. This will make the system more
robust and will feed the neural network more data, thus, boosting the
accuracy of the system. Training the TDNN and Bi-LSTM for more than
200 epochs is also a target since the computational limitation and
time constraints limited the total number of epochs to 10. Continuing
the training process for longer can further optimize the BCE objective
function leading to a better generalization capability of the system
on unseen data. We also plan to include more largescale datasets like
VoxCeleb~\cite{r-m:janin-2003} to further enhance the system.

\bibliographystyle{IEEEtran}
\bibliography{ms.bib}

\end{document}